\documentclass[twocolumn]{aastex631}

\usepackage{hyperref}

\graphicspath{{./}{Figures/}}

\begin{document}

\title{Calibrating the clock of {\em JWST}}

\author[0000-0002-8808-520X]{A. W. Shaw}
\affiliation{Department of Physics \& Astronomy, Butler University, 4600 Sunset Ave, Indianapolis, IN 46208, USA}

\correspondingauthor{A. W. Shaw}
\email{awshaw@butler.edu}

\author[0000-0001-6295-2881]{D. L. Kaplan}
\affiliation{Department of Physics, University of Wisconsin-Milwaukee, P.O. Box 413, Milwaukee, WI 53201, USA}


\author[0000-0003-3105-2615]{P. Gandhi}
\affiliation{School of Physics \& Astronomy, University of Southampton, Southampton, Hampshire SO17 1BJ, UK}

\author[0000-0003-0976-4755]{T. J. Maccarone}
\affiliation{Department of Physics and Astronomy, Texas Tech University, Lubbock, TX 79409, USA}


\author[0009-0002-6989-1019]{E. S. Borowski}
\affiliation{Department of Physics \& Astronomy, Louisiana State University, 202 Nicholson Hall, Baton Rouge, LA 70803, USA}

\author{C. T. Britt}
\affiliation{Space Telescope Science Institute, 3700 San Martin Drive, Baltimore, Maryland 21218, USA}

\author[0000-0002-7004-9956]{D. A. H. Buckley}
\affiliation{South African Astronomical Observatory, P.O Box 9, Observatory, 7935 Cape Town, South Africa}
\affiliation{Department of Astronomy, University of Cape Town, Private Bag X3, 7701 Rondebosch, South Africa}

\author[0000-0002-7226-836X]{K. B. Burdge}
\affiliation{Department of Physics, Massachusetts Institute of Technology, Cambridge, MA 02139, USA}
\affiliation{Kavli Institute for Astrophysics and Space Research, Massachusetts Institute of Technology, Cambridge, MA 02139, USA}

\author{P. A. Charles}
\affiliation{School of Physics \& Astronomy, University of Southampton, Southampton SO17 1BJ, UK}
\affiliation{Astrophysics, Department of Physics, University of Oxford, Denys Wilkinson Building, Keble Road, Oxford OX1 3RH, UK}
\affiliation{Department of Physics, University of the Free State, 205 Nelson Mandela Drive, Bloemfontein, 9300, South Africa}

\author[0000-0003-4236-9642]{V. S. Dhillon}
\affiliation{Department of Physics and Astronomy, University of Sheffield, Sheffield, S3 7RH, UK}
\affiliation{Instituto de Astrof{\'{\i}}sica de Canarias, E-38205 La Laguna, Tenerife, Spain}

\author[0000-0002-9858-9532]{R. G. French}
\affiliation{Space Science Institute, Boulder, CO, 80301 USA}

\author[0000-0003-3944-6109]{C. O. Heinke}
\affiliation{Department of Physics, University of Alberta, Edmonton, AB T6G 2E1, Canada}

\author[0000-0003-3318-0223]{R. I. Hynes}
\affiliation{Department of Physics \& Astronomy, Louisiana State University, 202 Nicholson Hall, Baton Rouge, LA 70803, USA}

\author[0000-0002-1116-2553]{C. Knigge}
\affiliation{School of Physics \& Astronomy, University of Southampton, Southampton, Hampshire SO17 1BJ, UK}

\author[0000-0001-7221-855X]{S. P. Littlefair}
\affiliation{Department of Physics and Astronomy, University of Sheffield, Sheffield, S3 7RH, UK}'

\author{Devraj Pawar}
\affiliation{R. J. College, Mumbai-86, India}

\author[0000-0002-7092-0326]{R. M. Plotkin}
\affiliation{Department of Physics, University of Nevada, Reno, NV 89557, USA}
\affiliation{Nevada Center for Astrophysics, University of Nevada, Las Vegas, NV 89154, USA}

\author[0000-0001-5644-8830]{M. E. Ressler}
\affiliation{Jet Propulsion Laboratory, California Institute of Technology, 4800 Oak Grove Drive, Pasadena, CA 91109, USA}

\author[0000-0002-1123-983X]{P. Santos-Sanz}
\affiliation{Instituto de Astrof{\'{\i}}sica de Andaluc{\'{\i}}a (CSIC), Glorieta de la Astronom{\'{\i}}a s/n, 18008-Granada, Spain}

\author[0000-0003-1331-5442]{T. Shahbaz}
\affiliation{Instituto de Astrof{\'{\i}}sica de Canarias, E-38205 La Laguna, Tenerife, Spain}
\affiliation{Departamento de  Astrof\'{\i}sica, Universidad de La Laguna (ULL),  E-38206 La Laguna, Tenerife, Spain}

\author[0000-0001-6682-916X]{G. R. Sivakoff}
\affiliation{Department of Physics, University of Alberta, Edmonton, AB T6G 2E1, Canada}

\author[0000-0002-5041-3079]{A. L. Stevens}
\affiliation{Michigan State University Museum, 409 W. Circle Drive, East Lansing, MI 48824} 



\begin{abstract}

{\em JWST}, despite not being designed to observe astrophysical phenomena that vary on rapid time scales, can be an unparalleled tool for such studies. If timing systematics can be controlled, {\em JWST} will be able to open up the sub-second infrared timescale regime. Rapid time-domain studies, such as lag measurements in accreting compact objects and Solar System stellar occultations, require both precise inter-frame timing and knowing when a time series begins to an absolute accuracy significantly below 1s. In this work we present two long-duration observations of the deeply eclipsing double white dwarf system ZTF\,J153932.16$+$502738.8, which we use as a natural timing calibrator to measure the absolute timing accuracy of {\em JWST}'s clock. From our two epochs, we measure an average clock accuracy of $0.12\pm0.06$s, implying that {\em JWST} can be used for sub-second time-resolution studies down to the $\sim100$ms level, a factor $\sim5$ improvement upon the pre-launch clock accuracy requirement. We also find an asymmetric eclipse profile in the F322W2 band, which we suggest has a physical origin.

\end{abstract}


\section{Introduction} 
\label{sec:intro}

One of the most fundamental aspects of observational astronomy is the measurement of flux variability on all timescales. From the Crab Pulsar's ns sub-pulses \citep{Hankins-2003} and the ms pulsations seen in the fastest rotating neutron stars \citep{Hessels-2006,Ambrosino-2017} to the slow, aperiodic variability seen in the multi-wavelength light curves of active galactic nuclei \citep{Vaughan-2003}, it is clear that variability is ubiquitous across all astrophysical phenomena. 

The sub-second regime, in particular, has emerged as a key area of study in multiple sub-fields of astrophysics. In accreting compact objects such as black holes and neutron stars, rapid stochastic variations are seen across the electromagnetic spectrum, and optical and infrared variations have been seen to lag behind X-ray emission on characteristic timescales as short as $\approx0.1$s \citep[e.g.][]{Gandhi-2008,Casella-2010,Gandhi-2017,Paice-2019}. Stellar occultations by Solar System minor bodies and their ring systems can last less than a second \citep[see e.g.][]{Braga-Ribas-2014,Ortiz-2017,Morgado-2023}, such that sub-second time resolution observations can help map kilometer-level features.\footnote{Indeed, a stellar occultation by the rings of the Centaur Chariklo was predicted and subsequently observed with {\em JWST} on October 18, 2022 demonstrating the power and applicability of this technique using {\em JWST} (Santos-Sanz et al., submitted).} Pulsars are precise cosmic clocks that show a wide variety of periodicities, from the accretion powered millisecond pulsars \citep[e.g.][]{Patruno-2021}, to the much slower ($\geq2$s) magnetars \citep[e.g.][]{Kaspi-2017}. Pulsars have been known to exhibit phase lags between wavelength bands on sub-second \citep[e.g.][]{Dhillon-2009} and even sub-millisecond time scales \citep[e.g.][]{Papitto-2019} that can allow us to identify the sites of particle acceleration.

In all of the aforementioned objects, the mid-infrared (MIR) remains a relatively under-studied wavelength regime, mostly due to the lack of available instrumentation that is both able to observe at micron wavelengths {\it and} do it on fast (sub-second) timescales. However with the successful December 2021 launch of {\em JWST}, we are in a new era of infrared astronomy. Despite not being designed for observing rapidly variable objects at high time-resolution, {\em JWST} can be an unparalleled tool for such studies - if timing systematics can be understood and controlled - owing to its ability to obtain high signal-to-noise, high time-resolution imaging and spectroscopy in the MIR. In order to open up the MIR to the sub-second timing community, we require knowledge of the absolute timing accuracy of {\em JWST}'s clock, which had a pre-launch accuracy requirement of $\sim0.5$s (K. Stevenson; priv. comm.). The goal of this work is to establish a high-precision validation of the in-orbit accuracy of the mission clock, including a check on the clock's drift. To do this we used {\em JWST} to observe a detached, eclipsing double white dwarf (WD) binary ZTF\,J153932.16$+$502738.8 \citep[hereafter, ZTF\,J1539;][]{Burdge-2019} as a natural timing calibrator. The use of a natural timing calibrator to calibrate the clock on a space mission has previously been demonstrated by \citet{Fortier-2024}, who used observations of the short-period eclipsing binary HW\,Vir to demonstrate a clock accuracy of better than 1s for the CHaracterising ExOPlanet Satellite \citep[{\em CHEOPS};][]{Benz-2021}.

ZTF\,J1539 consists of a hot ($\sim50,000$K) 0.61$M_{\rm \odot}$ CO-core WD primary and a cooler ($<10,000$K) 0.21$M_{\rm \odot}$ He-core WD secondary. The system has a very short orbital period ($P_{\rm orb}=6.91$min), shows deep primary and modest secondary eclipses and has a very well-known ephemeris calculated from ground-based observations \citep{Burdge-2019}, making it an ideal calibration source. In principle, the precision with which any given sampled light curve can be matched to a known ephemeris model is estimated as the uncertainty in determining the time of binary conjunction, $t_c$:

\begin{equation}
    \sigma_{t_c} \approx \frac{1}{Q}T\sqrt{\frac{\tau}{2T}}
    \label{eq:Winn}
\end{equation}

\noindent where $T$ represents the interval between the halfway points of ingress and egress, $\tau$ is the ingress or egress duration (i.e. time from first contact to second contact) and $Q=\sqrt{N}\delta/\sigma$. $N$ is the number of samples obtained during eclipse, $\delta$ represents the fractional eclipse depth, and $\sigma$ is the typical relative (un-eclipsed, baseline) flux uncertainty \citep{Winn-2010}.

We designed our observing plan with Equation \ref{eq:Winn} in mind. Based on the phase-folded light curves presented by \citet{Burdge-2019}, we found that we need to obtain $N\sim100$ data points during the eclipse to achieve a timing accuracy of $\sigma_{t_c} \sim 0.1$s, or $N\sim390$ for $\sigma_{t_c} \sim 0.05$s, which motivated a long-duration, high time-resolution observation of ZTF\,J1539 with {\em JWST}.

In Section \ref{sec:obs} we present our observations and data reduction, before describing our analysis and results in Section \ref{sec:analysis}. We discuss the results in Section \ref{sec:discussion} and summarize them in Section \ref{sec:conclusions}. We anticipate that results from this study will be useful for planning future high time-resolution observations of a wide variety of variable objects with {\em JWST}.

\section{Observations and Data Reduction} 
\label{sec:obs}

\begin{table*}
    \centering
    \caption{Summary of NIRCam observations for {\em JWST} proposal ID 1666}
    \begin{tabular}{l c c c c c}
    \hline
        Epoch & Start Time  & Total Exposure$^{\rm a}$  & Filter  &  Average Specific Intensity$^{\rm b}$ & S/N$^{\rm c}$\\
        &(UTC)&(s)&&(MJy sr$^{-1}$) & \\
        \hline\hline
         1 &  2023-02-05 19:02:07 & 32294.376 & F070W & $713.9\pm0.7$ & 12.4\\ 
         1 &  2023-02-05 19:02:07 & 32294.376 & F277W & $22.78\pm0.03$ & 11.5\\
         2 & 2023-07-18 18:52:09 & 32294.376 & F070W & $654.4\pm0.7$ & 11.5\\
         2 & 2023-07-18 18:52:09 & 32294.376 & F322W2 & $20.03\pm0.02$ & 17.1\\
         \hline 
    \end{tabular} \\
    \raggedright\footnotesize$^{\rm a}$ Total exposure time, distinct from the on-source time, i.e. the length of time that {\em JWST} was pointed at ZTF\,J1539 which was $9.45$h \\
    \raggedright\footnotesize$^{\rm b}$ Weighted average specific intensity in the non-eclipsed phases \\
    \raggedright\footnotesize$^{\rm c}$ Typical signal-to-noise ratio for a single integration in the non-eclipsed phases
    \label{tab:obs}
\end{table*}

We observed ZTF\,J1539 as part of PID 1666 (PI: Gandhi) using {\em JWST}'s Near Infrared Camera \citep[NIRCam;][]{Rieke-2023} in the SUB160P subarray mode, which provided a frame time of 0.27864s\footnote{\href{https://jwst-docs.stsci.edu/jwst-near-infrared-camera/nircam-instrumentation/nircam-detector-overview/nircam-detector-subarrays}{https://jwst-docs.stsci.edu/jwst-near-infrared-camera/nircam-instrumentation/nircam-detector-overview/nircam-detector-subarrays}}. We performed two observations, separated by approximately 5 months. The first observation, hereafter referred to as Epoch 1, commenced at 2023-02-05 19:02:07 UTC for an on-source time of 9.45h. We utilized the F070W filter (pivot wavelength, $\lambda_{\rm pivot}=0.704~\mu$m, bandwidth $\Delta\lambda=0.128~\mu$m)\footnote{\href{https://jwst-docs.stsci.edu/jwst-near-infrared-camera/nircam-instrumentation/nircam-filters}{https://jwst-docs.stsci.edu/jwst-near-infrared-camera/nircam-instrumentation/nircam-filters}} for NIRCam's short wavelength channel and the F277W filter ($\lambda_{\rm pivot}=2.786~\mu$m, $\Delta\lambda=0.672~\mu$m) in the long wavelength channel. The second observation, hereafter referred to as Epoch 2, commenced at 2023-07-18 18:51:38 UTC, for an on-source time of 9.45h. We once again utilized the F070W filter in the short wavelength channel, but opted for the F322W2 filter ($\lambda_{\rm pivot}=3.247~\mu$m, $\Delta\lambda=1.339~\mu$m) for the long wavelength channel. The broader bandpass of F322W2 resulted in a higher signal-to-noise ratio in the long wavelength channel. In both epochs and for both wavelength channels, an exposure consisted of 6100 integrations, with each integration consisting of 10 groups and each group containing one frame. We chose the BRIGHT1 readout pattern which gave an effective exposure time per integration of 5.29s.\footnote{\href{https://jwst-docs.stsci.edu/jwst-near-infrared-camera/nircam-instrumentation/nircam-detector-overview/nircam-detector-readout-patterns}{https://jwst-docs.stsci.edu/jwst-near-infrared-camera/nircam-instrumentation/nircam-detector-overview/nircam-detector-readout-patterns}} Observations are summarized in Table \ref{tab:obs}.

We downloaded the pipeline-processed (version 1.11.4) high-level data products from the Barbara A. Mikulski Archive for Space Telescopes (MAST)\footnote{\href{https://mast.stsci.edu}{https://mast.stsci.edu}}. We performed all of our data analysis on the background subtracted, calibrated, cosmic-ray flagged rate data, which exists as a data cube with the label {\tt crfints} produced by the {\em JWST} pipeline.\footnote{\href{https://jwst-pipeline.readthedocs.io/en/latest/jwst/data_products/science\_products.html\#}{https://jwst-pipeline.readthedocs.io/en/latest/jwst/\\data\_products/science\_products.html\#}} Each individual slice of the {\tt crfints} cubes contained an image of the ZTF\,J1539 field, with a $5\arcsec\times5\arcsec$ and $10\arcsec\times10\arcsec$ field of view for the short and long wavelength channels, respectively.

To extract precise fluxes of ZTF\,J1539 from each integration we performed point spread function (PSF) photometry using the {\tt photutils} package \citep{Bradley-2023}, an {\tt astropy}-affiliated {\sc python} package \citep{Astropy-2013,Astropy-2018,Astropy-2022}. We also utilized the {\tt WebbPSF} {\sc python} package \citep{Perrin-2012,Perrin-2014}, which generates model PSFs for all of {\em JWST}'s instruments dependent on input parameters such as the subarray mode, detector, source position on the detector and the filter. 

We first combined all the images to make an average image for each wavelength channel. We then used the {\tt centroid\_sources} function to fit the mean centroid of the point source, relative to the origin of the SUB160P subarray. ZTF\,J1539 was the only visible source in the field-of-view in both wavelength channels. We then used {\tt WebbPSF} to generate a model PSF at this average source position, again for each wavelength channel. No spectral information was provided to {\tt WebbPSF} and thus the resultant PSF was monochromatic.

\begin{figure*}
    \centering
    \includegraphics[width=0.95\textwidth]{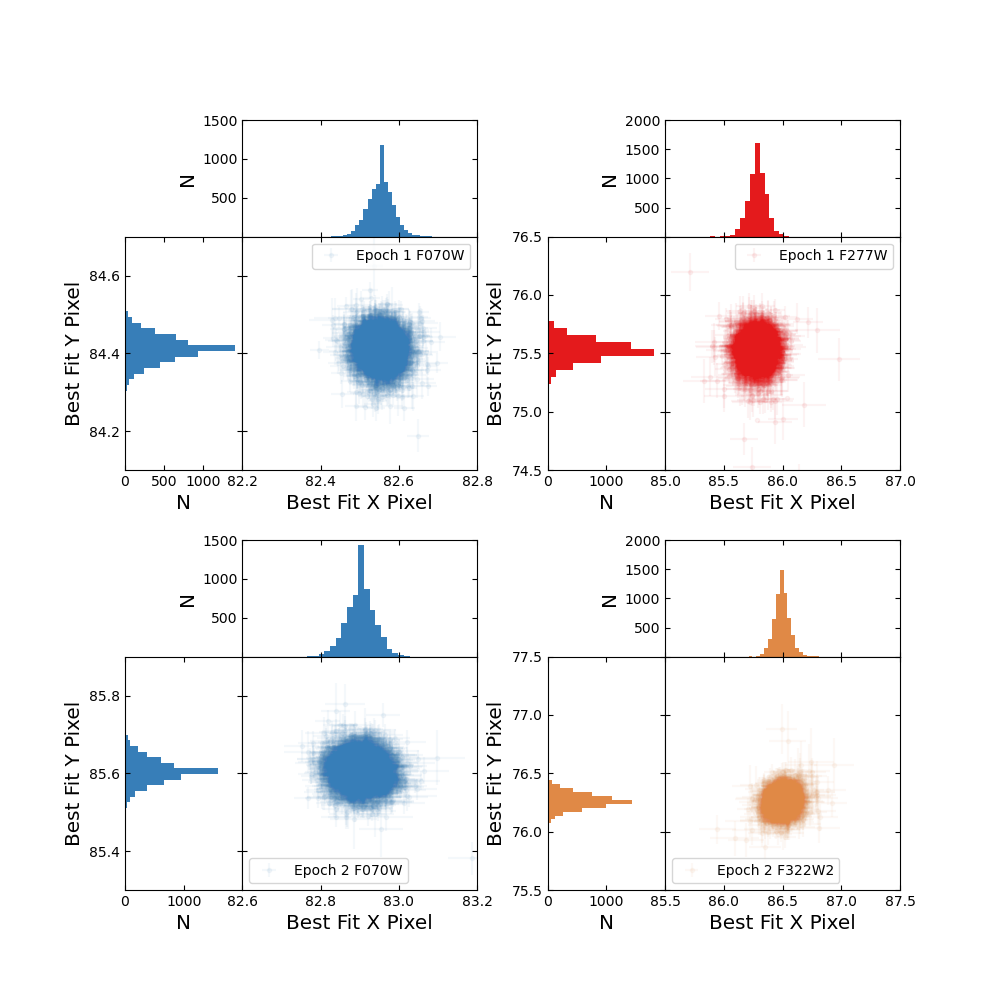}
    \caption{Best-fit position of ZTF\,J1539 on the {\em JWST} NIRCam detector in the short (left) and long wavelength channels for Epochs 1 (top) and 2 (bottom). X and Y pixel positions were determined by the {\tt PSFPhotometry} function in {\tt photutils}. In each panel we also show the 1-dimensional histograms of best-fit X and Y pixel values, where $N$ represents the number of pixels in each histogram bin.}
    \label{fig:pos}
\end{figure*}

With the model PSF generated, we then performed basic aperture photometry on each image at the location of the source, as determined by the centroiding process described above. We used a circular aperture of radius 2 pixels ($0.061\arcsec$) and 3 pixels ($0.188\arcsec$) for the short and long wavelength channels, respectively. The background was measured from an annulus centered on the source, with inner and outer radii of 10 and 20 pixels, respectively. The background-subtracted flux from the aperture photometry, along with the source centroid determined from the average image, were used as first guesses for the {\tt PSFPhotometry} function, which calculated a best-fit position and flux for ZTF\,J1539 for every individual exposure. Fig. \ref{fig:pos} shows the best-fit positions in pixel space for each integration, showing that the change in centroid between the majority of images is on the sub-pixel level.

As ZTF\,J1539 is an eclipsing source, we had to consider how we measured flux in frames when the source was not visible. For images where the signal-to-noise ratio of the source was $<3$ in the short wavelength channel or $<2$ in the long wavelength channel -- the limit at which the centroiding became unreliable in each channel -- we fixed the location of the PSF to the average centroid. Again, using the background subtracted flux from the aperture photometry as an initial flux, we used {\tt PSFPhotometry} to fit the flux at this fixed location, which, at totality, is essentially equivalent to the noise level of the data. Flux uncertainties were extracted using the ERR extension of the pipeline-processed data file, which is calculated from the Poisson and read noise in each detector pixel.\footnote{We note that, for the short waveband detector, the ERR extension includes a conservative estimate of pre-amplifier reset noise such that errors are possibly slightly overestimated. This is owing to the fact that the SUB160P subarray has no reference pixels on the full NRCB1 detector ({\em JWST} Help Desk Ticket INC0190875).} The raw, unfolded light curves are presented in Figs. \ref{fig:E1_raw_lc} and \ref{fig:E2_raw_lc}, highlighting the quality of the data and the length of the observation.

\section{Analysis and Results}
\label{sec:analysis}

\subsection{Timing Analysis}
\label{sec:timing}

{\em JWST} has two clocks on board. One is on the Integrated Science Instrument Module (ISIM), where all of {\em JWST}'s instruments are housed, and the second is the spacecraft clock. The `INT\_TIMES' FITS extension, from which we derive the light curve time stamps, is calculated from the ISIM flight software (B. Hilbert, priv. comm.)\footnote{{\em JWST} Help Desk Ticket INC0196358}. The ISIM clock is updated from the spacecraft clock every 64 ms, and the accuracy of the spacecraft clock is checked and updated from the ground during every ground station contact (K. Stevenson, priv. comm.). The time between ground station contacts varies and depends on what other missions are underway in that direction at the same time, but they generally last between 2 and 6 hours and generally happen 2-3 times per day.\footnote{\href{https://blogs.nasa.gov/webb/2023/08/15/talking-with-webb-using-the-deep-space-network/}{https://blogs.nasa.gov/webb/2023/08/15/talking-with-webb-using-the-deep-space-network/}}

The spacecraft clock is known to drift, and the drift is required to be kept to less than 42 ms per day ($<0.4 \mu{\rm s}~{\rm s}^{-1}$; B. Hilbert, priv. comm.). Finally, the time stamps that are assigned to the INT\_TIMES FITS extension may vary from actual end of the readout time due to delays of up to $\sim10$ ms in the transfer of the readout data from the detector through the ISIM to the solid-state recorder (B. Hilbert, priv. comm.). All these sources of error on the clock imply a maximum difference of $\approx116$ms between the published time stamps and reality.

In order to quantify the accuracy of {\em JWST}'s clock and compare it to this expected value, we folded the NIRCam data on the latest ephemeris for ZTF\,J1539, as calculated from ground-based data, defining phase = 0 as the middle of the primary eclipse. We assumed  the following:

\begin{center}
    $T_0=59341.97656626(26)$ BMJD$_{\rm TDB}$
\end{center}

\begin{center}
    $P_{\rm orb}=0.0048008034992(13)$ d
\end{center}

\begin{center}
    $\dot{P}_{\rm orb}=-2.36812(44)\times10^{-11}$ s s$^{-1}$
\end{center}

\noindent where $T_0$ represents the mid-eclipse time of the primary eclipse, $P_{\rm orb}$ is the binary orbital period and $\dot{P}_{\rm orb}$ is the period derivative (K. Burdge, priv. comm.). We note that this ephemeris is an update to the one published by \citet{Burdge-2019} and this work marks the first publication of these values. Information on the observations behind this updated ephemeris is provided in Section \ref{sec:E2}.

As discussed above, light curve time stamps are derived from the `INT\_TIMES' FITS extension. Specifically, we utilize the `int\_mid\_BJD\_TDB' column, which contains the mid-time of each integration, corrected to the Solar System's barycenter and on to the Barycentric Dynamical Time (TDB) standard. The barycentric correction is performed during the level 1b stage of the standard pipeline processing. Relevant to the barycentric correction, the position of JWST is measured by the Deep Space Network during every ground contact and the maximum uncertainty on JWST's position and velocity is 50km and 20 mm s$^{-1}$, respectively.\footnote{\href{https://ntrs.nasa.gov/citations/20220010187}{https://ntrs.nasa.gov/citations/20220010187}} This can result in sub-ms temporal uncertainties in the barycentric correction, but this is far smaller than the time scales we probe here.
 
We then used the {\tt ellc} package \citep{Maxted-2016} to model the phase-folded light curves and fit for the offset between $T_0$ and the mid-eclipse times as measured by {\em JWST}. We separated the fits by epoch, first characterizing the offset in Epoch 1, then in Epoch 2, but fitting the light curves from both bands in each epoch simultaneously. The binary parameters of ZTF\,J1539 are well-known, so we fixed the radii of the binary components, binary inclination and the mass ratio to the published values \citep{Burdge-2019}. In addition, we do not examine eccentricity or Doppler boosting in the light curves. We anticipate that these NIRCam data will be extremely useful in refining the binary parameters, as well as the physical characteristics of the binary components, but such work is beyond the scope of this study.
 
In modeling the {\em JWST}/NIRCam observations of ZTF\,J1539, we adopt the following set of wavelength-dependent free parameters: the surface brightness ratio of the secondary's unheated face to that of the primary ($J$), a heating parameter to fit the sinusoidal variation of the light curve due to the irradiation of the secondary (heat$_2$), a simple linear limb-darkening model \citep{Claret-2011,Gianninas-2013,Claret-2020}, with coefficients for the primary (ldc$_1$) and secondary (ldc$_2$), and a gravity-darkening coefficient for the secondary (gdc$_2$). \citet{Burdge-2019} characterize these parameters of ZTF\,J1539 with $g'$-band observations. However, we are utilizing observations in the red (F070W) and infrared (F277W/F322W2) bands, and thus must fit for new values of $J$, heat$_2$, ldc$_1$, ldc$_2$ and gdc$_2$ for each NIRCam filter used to observe ZTF\,J1539. In the context of this work, the most important parameter in our parameter search is the wavelength-independent offset between the mid-eclipse time as measured by {\em JWST}, $T_{0, {\rm JWST}}$, and the known $T_0$ that we folded the {\em JWST} data on, which we label $\Delta T_0$. This is the parameter that will allow us to derive the accuracy of {\em JWST}'s clock.

We utilize the {\tt emcee} {\sc python} package \citep{Foreman-Mackey-2013} to perform a Markov Chain Monte Carlo (MCMC) parameter search. We initialize 50 walkers in a parameter space around reasonable first estimates of the free parameters \citep{Claret-2011,Gianninas-2013,Burdge-2019,Claret-2020} and iterate for a maximum of 100,000 steps, with a burn-in of a number of steps equal to double the maximum autocorrelation time $\tau$ for all free parameters. We fit both the short- and long-wavelength channel light curves simultaneously, as light curve time stamps are derived from the same internal clock for both channels. We monitor $\tau$ and consider the fit to be converged if the chain is longer than $100\tau$ for all free parameters.

In Epoch 1, we find $\Delta T_0=0.08\pm0.07$s, where the uncertainties are derived from the $1\sigma$ confidence interval of the $\Delta T_0$ posterior probability distribution. We can compare this uncertainty to those predicted by Equation \ref{eq:Winn}. In each band we measure a total of $N=820$ flux points during the eclipses, the fractional uncertainties $\sigma_f$ can be derived from the signal-to-noise ratios in Table \ref{tab:obs} and we measure an eclipse depth of $\delta=0.67$ in F070W and $\delta=0.49$ in F277W, relative to the mean non-eclipsed flux. Using Equation \ref{eq:Winn} we would expect to measure $\sigma_{t_c} \approx 0.08$ in the F070W band and $\sigma_{t_c} \approx 0.12$ in the F277W band which, when combined, implies a total uncertainty of $\sigma_{t_c} \approx 0.07$, consistent with what we measure.

\begin{figure*}
    \centering
    \includegraphics[width=\textwidth]{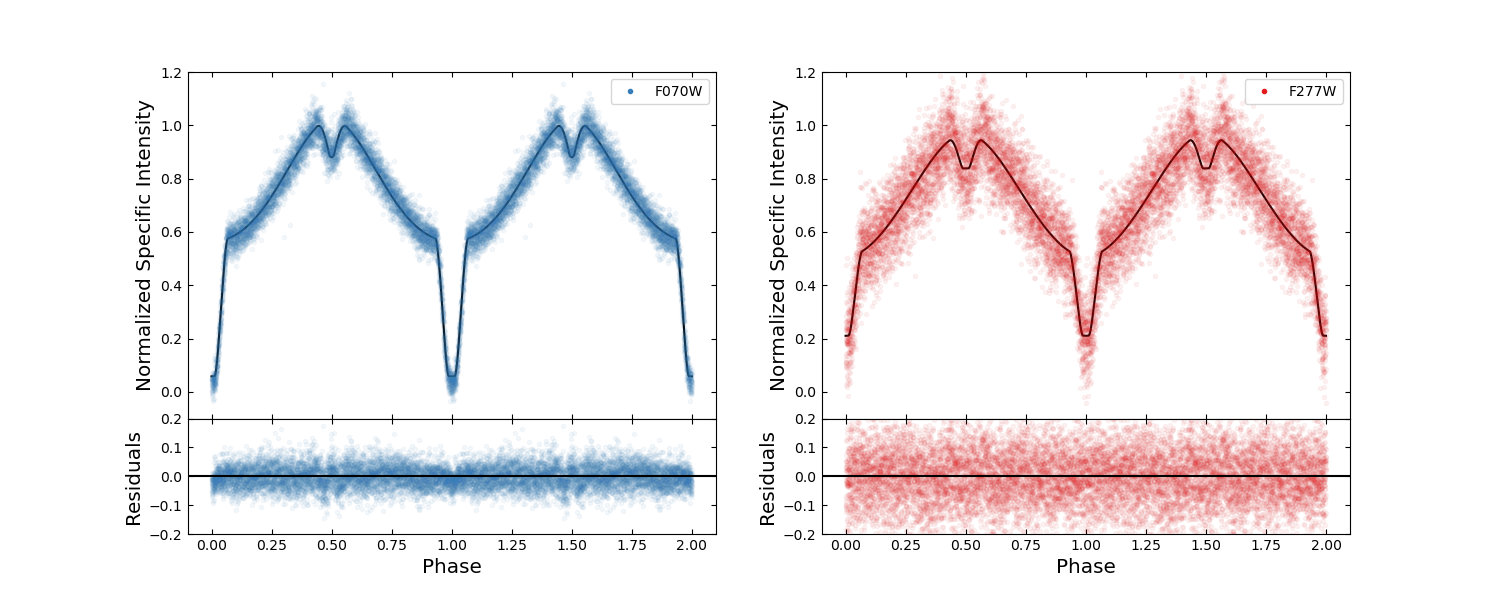}
    \caption{{\em JWST}/NIRCam Epoch 1 light curves of ZTF\,J1539 in the F070W (left) and F277W (right) bands, folded on the most recent ephemeris. In both sub-figures we also show the best-fit model to the light curve as a solid line and the resultant residuals in the bottom panels of each sub-figure.}
    \label{fig:E1_lcfit}
\end{figure*}

We show the full corner plots derived from this fit in Fig. \ref{fig:E1_corner}, noting that the most important parameter for this study, $\Delta T_0$, is not dependent on our choice of limb-darkening law. We show the folded light curves and best-fit model in Fig. \ref{fig:E1_lcfit}.

\begin{figure*}
    \centering
    \includegraphics[width=\textwidth]{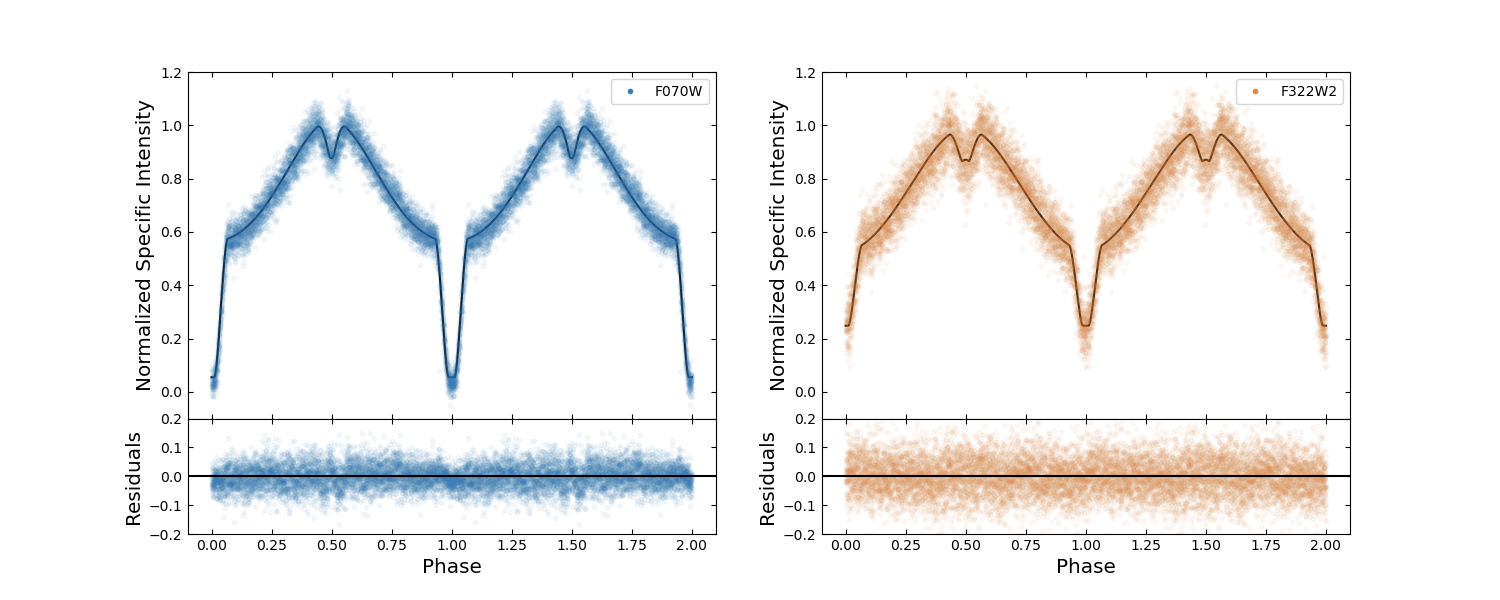}
    \caption{{\em JWST}/NIRCam Epoch 2 light curves of ZTF\,J1539 in the F070W (left) and F322W2 (right) bands, folded on the most recent ephemeris. In both sub-figures we also show the best-fit model to the light curve as a solid line and the resultant residuals in the bottom panels of each sub-figure.}
    \label{fig:E2_lcfit}
\end{figure*}

In Epoch 2, we find a much larger timing offset, measuring $\Delta T_0=-0.25\pm0.07$s (again, with an uncertainty consistent with what is expected from Equation \ref{eq:Winn}). We show the full corner plots derived from this fit in Fig. \ref{fig:E2_corner} and the folded light curves in Fig. \ref{fig:E2_lcfit}.

\section{Discussion}
\label{sec:discussion}

\subsection{Further investigation of Epoch 2}
\label{sec:E2}

The large $\Delta T_0$ in Epoch 2 may raise questions about the accuracy of the ephemeris on which we folded the light curves. However, the ephemeris provided by K. Burdge (priv. comm.) has been calculated based on six years of regular ground-based optical monitoring of ZTF\,J1539. Data used in the calculation include observations from the High PERformance CAMera (HiPERCAM) on the 10.4m Gran Telescopio Canarias in La Palma, Spain, which has an absolute timing accuracy better than $\sim100~\mu{\rm s}$ \citep{Dhillon-2021}, as well as the Caltech HIgh-speed Multi-colour camERA (CHIMERA) on the 200 inch Hale telescope at Palomar Observatory, USA, which has an absolute timing accuracy of $\sim1$ms \citep{Harding-2016}. In addition, the uncertainties on the reported ephemeris are extremely small (see Section \ref{sec:timing}), with an uncertainty on $T_0$ of just 0.022s. It seems unlikely, then, that the measured $\Delta T_0$ in Epoch 2 is due to inaccuracy in the ephemeris.

Nevertheless, for a secondary check, we utilized ground-based observations of ZTF\,J1539 with the high-speed, quintuple-beam camera HiPERCAM that took place on 2023-07-26 22:21:15 UTC, just 8 days after {\em JWST} Epoch 2. Given that the absolute timing accuracy of HiPERCAM is better than $\sim100~\mu{\rm s}$ as discussed above, comparing the {\em JWST} Epoch 2 light curves to those from HiPERCAM's five filters ($u_sg_sr_si_sz_s$) can be seen as a confirmation of our measured offset from the ephemeris.

We observed ZTF\,J1539 on 2023-07-26 for 42.5\,min. A total of 1627 images were obtained simultaneously in $u_sg_sr_si_sz_s$ using exposure times of 1.560\,s with 0.008\,s dead time between each frame, where each HiPERCAM frame is GPS time-stamped to a relative (i.e. frame-to-frame) accuracy of 0.1$\mu$s and an absolute accuracy of 0.1\,ms \citep{Dhillon-2021}. HiPERCAM was used in full frame mode with 3$\times3$ binning, with the $u_s$ and $z_s$ CCDs set to skip two out of every three readouts, resulting in exposure times of 4.336\,s in these two filters.

The data were reduced using the HiPERCAM data reduction pipeline \citep{Dhillon-2021}. All frames were debiased and then flat-fielded, with the latter achieved by using the median of twilight-sky frames taken with the telescope spiralling. The CCD fringing pattern was removed from the $z_s$ HiPERCAM frames using the median of night-sky frames taken with the telescope spiralling. We used optimal photometry \citep{Naylor-1998} with software apertures that scaled in size with the seeing to extract the counts from ZTF\,J1539 and a bright comparison star in the same field of view, the latter acting as the reference for the PSF fits, transparency, and extinction corrections. The aperture position of ZTF\,J1539 relative to the comparison star was determined from the sum of 100 images, and this offset was then held fixed during the reduction so as to avoid aperture centroiding problems during primary eclipse. The sky level was determined from a clipped mean of the counts in an annulus surrounding each star and subtracted from the object counts.

\begin{figure}
    \centering
    \includegraphics[width=0.45\textwidth]{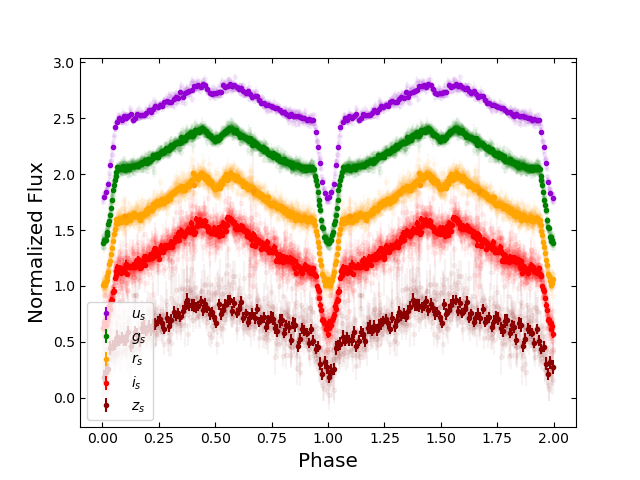}
    \caption{HiPERCAM light curves of ZTF\,J1539 in the $u_s$,$g_s$,$r_s$,$i_s$ and $z_s$ bands, folded on the latest ephemeris. The opaque data points represent the binned data, while the raw data are shown as semi-transparent data points for each corresponding color}. Light curves have been normalized and offset from one another for clarity.
    \label{fig:HCAM_lcs}
\end{figure}

We corrected the HiPERCAM time stamps to the solar system barycenter, utilizing the {\tt astropy.time} package and the latest JPL ephemerides. This ensured that the HiPERCAM data are corrected to the same time system, TDB, as the {\em JWST} data. We then folded the HiPERCAM light curves on the latest ephemeris, as we did with the {\em JWST} light curves (Fig. \ref{fig:HCAM_lcs}) and performed a cross correlation between light curves from each HiPERCAM filter and each NIRCam filter. We calculated the cross-correlations using the z-transformed discrete correlation function \citep[ZDCF;][]{Alexander-1997,Alexander-2013}, which is designed for unevenly sampled light curves and comparing two light curves with different time stamps.

When combining all of the resultant ZDCFs, we measure the `time lag,' (which, in reality, is $\Delta T_{0, {\rm JWST}} - \Delta T_{0, {\rm HiPERCAM}}$) to be $-0.19\pm0.03$s, consistent with the measured timing offset of the {\em JWST} Epoch 2 light curves when compared directly with the most recent ephemeris for ZTF\,J1539.

As we discuss in Section \ref{sec:timing}, the {\it maximum} offset between the NIRCam time stamps and the true time is expected to be $\approx116$ms when accounting for all expected sources of errors in the clock, so the discrepancies in Epoch 2 are puzzling. To investigate further, we performed another MCMC parameter search on the phase-folded F070W and F322W2 light curves, but this time we assumed a separate $\Delta T_0$ for each waveband. Interestingly, we found that $\Delta T_0$ was not consistent between the two filters, despite time stamps originating from the same clock in both. In the F070W band we find $\Delta T_0=0.18\pm0.09$s, while in the F322W2 band we find $\Delta T_0=-0.78\pm0.10$s, meaning that the F322W2 observations are heavily biasing the overall measured offset in Epoch 2. We note that $\Delta T_0$ as measured from F070W alone, while not consistent with 0, is consistent with the maximum expected offset of $\approx116$ms. For completeness, we note here that we find $\Delta T_0=0.08\pm0.08$s and $\Delta T_0=0.08\pm0.13$s for the F070W and F277W light curves, respectively, in Epoch 1. Note also that the derived uncertainties are consistent with what is expected from Equation \ref{eq:Winn} (see Section \ref{sec:timing}).

\begin{figure}
    \centering
    \includegraphics[width=0.45\textwidth]{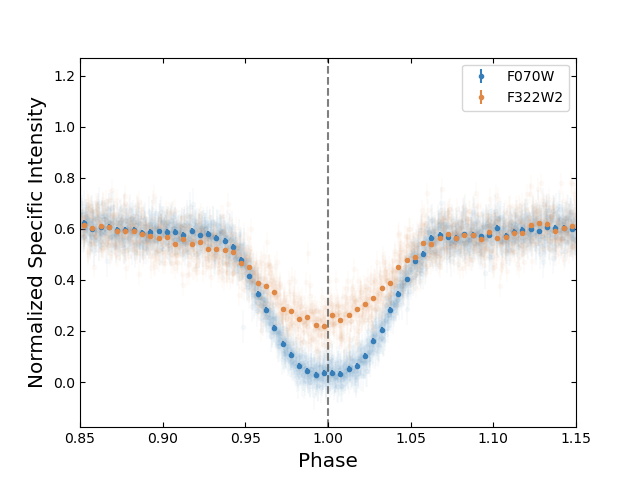}
    \caption{{\em JWST}/NIRCam light curves of ZTF\,J1539 in the F070W (blue) and F322W2 (orange) bands, folded on the most recent ephemeris and zoomed in on the primary eclipse.}
    \label{fig:E2_eclipse}
\end{figure}

The timing discrepancy between the two bands in Epoch 2 raises the possibility that it is in fact a physical characteristic of the system. In Fig. \ref{fig:E2_eclipse} we show the phase folded light curves from Epoch 2 in both bands, zoomed in on the primary eclipse. Compared to the F070W light curve, the F322W2 light curve is apparently asymmetric about the eclipse, with a slower rise out of the eclipse itself. It also appears that the ingress/egress in F322W2 is not symmetrical, which is clearer in the binned light curve. This is likely the source of the large $\Delta T_0$ offset in the F322W2 band, bearing in mind that $T_0$ is defined as the mid-eclipse time of the primary eclipse.

\begin{figure*}
\gridline{\fig{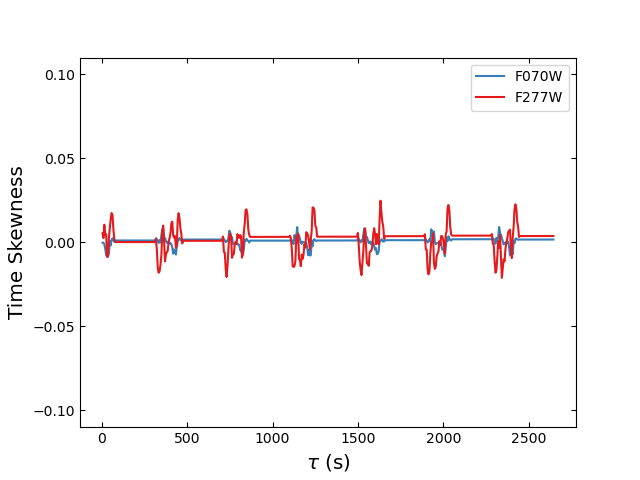}{0.5\textwidth}{}
          \fig{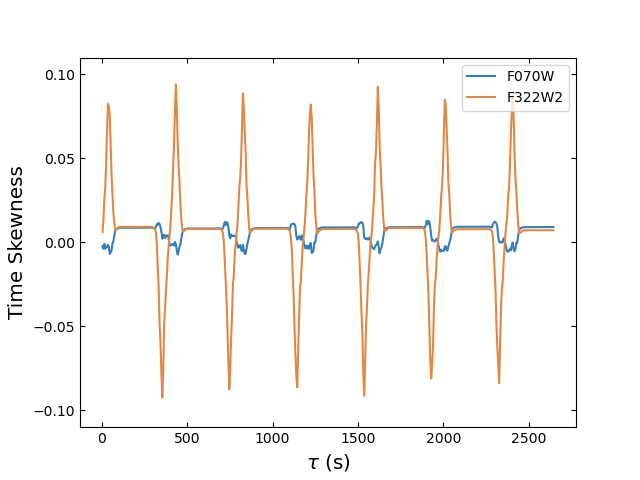}{0.5\textwidth}{}}
\caption{Time skewness of the {\em JWST}/NIRCam Epoch 1 (left) and Epoch 2 (right) light curves in the F070W (blue), F277W (red) and F322W2 (orange) bands.}
\label{fig:TS}
\end{figure*}

To confirm this, we followed the methodology of \citet{Maccarone-2002} who defined the time skewness (TS) parameter to measure the asymmetry of a light curve, based on a previous formulation by \citet{Priedhorsky-1979}. It is defined as follows:

\begin{equation}
    TS(\tau) = \frac{1}{\sigma^3}\overline{\left[ (s_i - \bar{s})^2(s_{i-u} - \bar{s}) - (s_i - \bar{s})(s_{i-u} - \bar{s})^2 \right]}
\end{equation}

\noindent where $\tau$ is defined to be $u$ times the time bin size (the time bin size is 5.29s for our light curves), $s_i$ is the flux of the $i$th element of the light curve, and $\sigma$ is the standard deviation of the flux. We only consider the primary eclipses in the light curve, as this is how $T_0$ is defined. We calculate the TS statistic for the F070W and F322W2 light curves from Epoch 2 and show the resultant plot in Fig. \ref{fig:TS}, alongside a plot of the TS statistic for Epoch 1 for comparison. We see that the TS for the F322W2 light curve is much larger than for the other light curves, indicating a strong asymmetry in the primary eclipse in this band. We plot the TS for values of $\tau$ equivalent to multiple orbital cycles of ZTF\,J1539 to show that the structure is consistent and not a one-off, implying that the skewness is intrinsic to the light curve rather than due to bad data (Fig. \ref{fig:TS}).

It is this asymmetric nature of the F322W2 light curve that has led us to conclude that the timing discrepancies in Epoch 2 have a physical origin. We have verified that the problem has not arisen from our photometry pipeline as there are no gaps in the light curve, and a real flux has been extracted from the source in every single image frame, eliminating the possibility of the pipeline filling a gap with an incorrect flux and potentially biasing the resultant light curves. The methodology by which we extract fluxes from the {\em JWST} images is consistent across all epochs and all wavebands, so it seems unlikely that this is a data analysis issue.

Some potential sources of the aysmmetry may include starspots on the surface of the secondary \citep[e.g.][]{Brinkworth-2005,Kilic-2015}, 
or perhaps a cyclotron feature that is present in the F322W2 band due to the presence of a strong magnetic field in one of the white dwarfs. We can speculate on the nature of the asymmetry in the infrared band, but will not investigate it further as the focus of this work remains on the accuracy of {\em JWST}'s clock. We therefore choose to proceed in our analysis by only considering the F070W light curve for Epoch 2, though we will utilize both the F070W and F277W light curves in Epoch 1 owing to the consistency of $\Delta T_0$ between the two bands. However, it is clear that our findings motivate a future spectral study of ZTF\,J1539, particularly at wavelengths $\lambda\gtrsim3~\mu$m.

\begin{figure}
    \centering
    \includegraphics[width=0.45\textwidth]{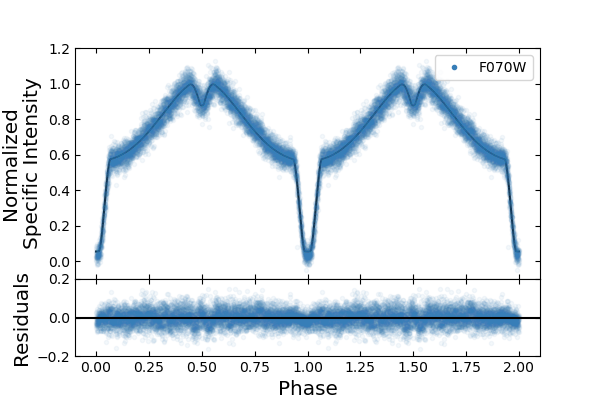}
    \caption{{\em JWST}/NIRCam Epoch 2 light curve of ZTF\,J1539 in the F070W band only, folded on the most recent ephemeris. We also show the best-fit model to the light curve -- after discarding the F322W2 light curve from the fit -- as a solid line, and the resultant residuals in the bottom panel.}
    \label{fig:E2_lcfit_SW}
\end{figure}

We therefore revise our measured offset to $\Delta T_0=0.18\pm0.09$s for Epoch 2. We include a new corner plot for Epoch 2 in the Appendix (Fig. \ref{fig:E2_corner_SW}), using only the F070W light curve for the MCMC parameter search. We also show the phase-folded light curve, along with the new best-fit model, in Fig. \ref{fig:E2_lcfit_SW}, and present our final table of best-fit parameters in Table \ref{tab:final_par}.

\begin{table}
    \centering
    \caption{Best-fit parameters to the phase-folded {\em JWST}/NIRCam light curves of ZTF\,J1539.}
    \begin{tabular}{l c c}
    \hline
         Parameter & Epoch 1 & Epoch 2  \\
    \hline \hline
        $\Delta T_0$ (s) &  $0.08\pm0.07$ & $0.18\pm0.09$ \\
        $J_{\rm F070W}$ & $0.015\pm0.001$ & $0.013\pm0.001$ \\
        heat$_{2,{\rm F070W}}$ & $6.15\pm0.09$ & $6.09^{+0.10}_{-0.09}$ \\
        ldc$_{1,\rm F070W}$ & $0.42^{+0.06}_{-0.11}$ & $0.44^{+0.05}_{-0.09}$ \\
        ldc$_{2,\rm F070W}$ & $1.42\pm0.02$ & $1.44\pm0.02$ \\
        gdc$_{2,\rm F070W}$ & $3.45\pm0.21$ & $3.54\pm0.23$ \\
        $J_{\rm F277W}$ & $0.122\pm0.003$ & -- \\
        heat$_{2,{\rm F277W}}$ & $11.52^{+0.25}_{-0.24}$ & --  \\
        ldc$_{1,\rm F277W}$ & $0.15^{+0.19}_{-0.11}$ & -- \\
        ldc$_{2,\rm F277W}$ & $0.85\pm0.04$ & -- \\
        gdc$_{2,\rm F277W}$ & $2.84^{+0.17}_{-0.16}$ & -- \\
    \end{tabular}
    \label{tab:final_par}\\
    \raggedright\footnotesize$\Delta T_0$: offset between the mid-eclipse time as measured by {\em JWST} and the known mid-eclipse time from the ephemeris.\\
    \raggedright\footnotesize$J$: surface brightness ratio of the unheated face of the secondary to that of the primary. Filter dependent.\\
    \raggedright\footnotesize${\rm heat}_2$: a heating parameter describing the irradiation of the secondary. Filter dependent.\\
    \raggedright\footnotesize${\rm ldc}_{1/2}$: limb-darkening coefficient for the primary (1) and the secondary (2). Filter dependent.\\
    \raggedright\footnotesize${\rm gdc}_{2}$: gravity-darkening coefficient for the secondary (2). Filter dependent.

\end{table} 

\subsection{Clock Drift}
\label{sec:drift}

\begin{figure}
    \centering
    \includegraphics[width=0.5\textwidth]{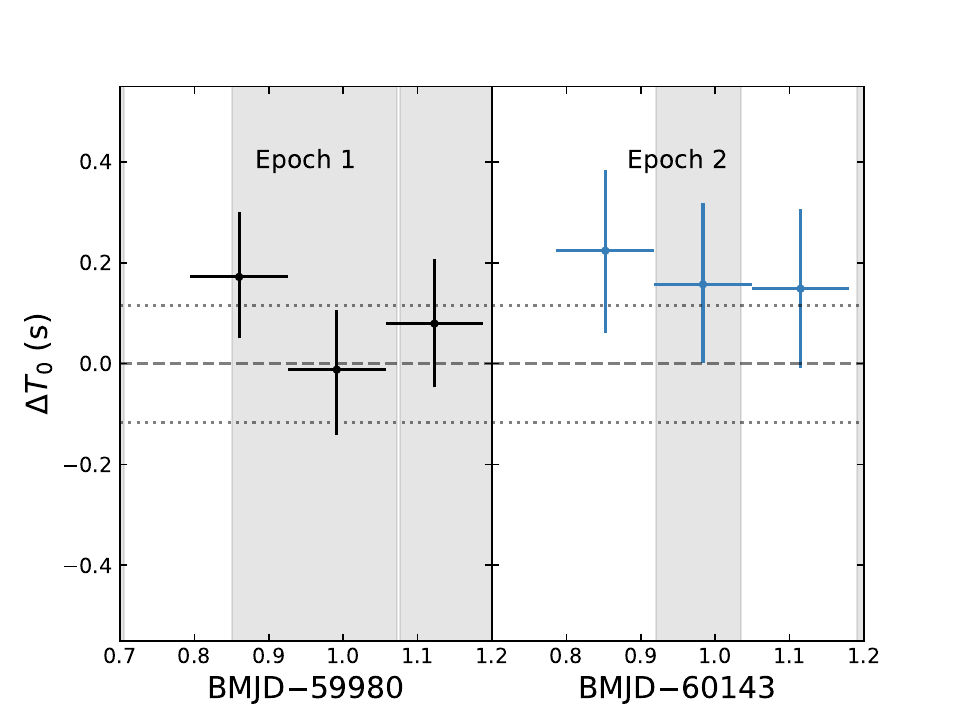}
    \caption{The evolution of $\Delta T_0$ with time, tracking the drift of {\em JWST}'s clock. In Epoch 2, the points are colored blue to represent the fact that we only used the F070W data to measure the drift. The grey shaded regions represent the times at which {\em JWST} was in contact with a ground station, at which time the spacecraft clock is updated. The horizontal dotted lines in each panel represent the maximum expected offset of $\Delta T_0\approx116$ms when accounting for all expected sources of errors in the clock (see Section \ref{sec:timing}).}
    \label{fig:drift}
\end{figure}

As we discuss in Section \ref{sec:timing}, {\em JWST}'s clock is expected to drift between ground station contacts for a maximum of 42 ms every 24 hours. We test the clock drift with our observations by splitting each epoch into segments and measuring the evolution, if any, of $\Delta T_0$ across the segments. To do this, we fix every parameter to their best-fit values from the fit to the entire light curve, except for $\Delta T_0$. We then split the light curves from each epoch into three segments of $\sim3$ h each and performed an MCMC parameter search for $\Delta T_0$, fitting both bands at the same time in Epoch 1 and F070W only in Epoch 2. We plot the results in Fig. \ref{fig:drift}, where we also include the times where {\em JWST} was in contact with a ground station\footnote{{\em JWST} Help Desk Ticket INC0194752} and the clock was corrected.

In both epochs, we see that all of the light curve segments have a $\Delta T_0$ that is completely consistent with the maximum expected offset of $\approx116$ms, given the known sources of uncertainty in the clock, which includes the drift. While there appears to be a shift in $\Delta T_0$ in Epoch 1, coincident with a ground station contact, this is not statistically significant. We are therefore unable to confirm any improvement in the accuracy of {\em JWST}'s clock after contact with a ground station.

\subsection{Relative Timing Precision}
\label{sec:relclock}

We have also calculated the {\em relative} timing precision of {\em JWST}'s clock by measuring the time intervals between each of the 6100 integrations for each set of observations. For all bands and epochs, we found that the exposure times remained constant to better than 0.6 $\mu$s.

\section{Conclusions}
\label{sec:conclusions}

We have used {\em JWST}/NIRCam to observe the eclipsing double white dwarf binary ZTF\,J153932.16+502738.8, which has a very well-characterized ephemeris. We compared the known mid-point of the primary eclipse with the one we measured with {\em JWST} in order to derive the absolute accuracy of {\em JWST}'s clock. When comparing the ephemeris with the first epoch of {\em JWST}/NIRCam data, we find $\Delta T_0=0.08\pm0.07$s, where this value was derived by fitting a model to both the F070W and F277W light curves. In Epoch 2 we used an ultra-wide filter for the long wavelength channel, centered at a slightly longer wavelength than in Epoch 1. We found that the resultant F322W2 light curve showed an asymmetric primary eclipse, which we suggest has a physical origin. As a result, we derive $\Delta T_0=0.18\pm0.09$s for Epoch 2 using only the F070W light curve. 

Considering that the accuracy of the ephemeris of ZTF\,J1539 is well-documented, we associate our measured offsets with the true absolute timing accuracy of {\em JWST}'s onboard clock. A weighted average of the two epochs implies a clock accuracy of $0.12\pm0.06$s. 


Discussions with the {\em JWST} Help Desk have helped us identify the main sources of error in the time stamps that are recorded in the {\em JWST} INT\_TIMES FITS extension, and these errors can account for a maximum $\approx116$ ms difference (see section \ref{sec:timing}). This is consistent with what we measure in both epochs (albeit by eliminating the F322W2 from our analysis in Epoch 2) and implies that sub-second absolute timing is possible with {\em JWST}, down to the $\sim100$ms level. Our results significantly improve upon the pre-launch clock accuracy requirement of $\sim0.5$s (K. Stevenson; priv. comm.).

With respect to the planning of future observations with {\em JWST} with a need for sub-second timing accuracy, we note that the $\sim100$ms accuracy level that we present here is relevant for all instruments on board the spacecraft. This opens up the sub-second timing regime to, for example, the mid-IR (both imaging and spectroscopy) via the Mid-Infrared Instrument \citep[MIRI;][]{Rieke-2015a,Rieke-2015b}. The mid-infrared is poorly explored in the sub-second regime, but with knowledge of the precision of {\em JWST}'s clock, we can start to probe this wavelength range on fast time-scales. Work on this has already begun in the field of black hole X-ray binaries \citep{Gandhi-2024}.

\section*{Acknowledgments}
We dedicate this paper to the memory of our dear colleague Prof. Tom Marsh, who is greatly missed. We credit Tom with the selection of ZTF J153932.16$+$502738.8 as a timing calibrator for this program.

We thank the anonymous referee for feedback that helped improve the manuscript. AWS thanks the JWST time series observations working group, who provided useful feedback on our results. AWS and DLK acknowledge funding from the Space Telescope Science Institute, STScI, operated by AURA under grant number JWST-GO-01666.022-A. PG acknowledges funding from the Royal Society (SRF$\backslash$R1$\backslash$241074) and UKRI Science \& Technology Facilities Council. COH acknowledges support from NSERC Discovery Grant RGPIN-2023-04264. PS-S acknowledges ﬁnancial support from the Spanish I+D+i Project PID2022-139555NB-I00 (TNO-JWST) and the Severo Ochoa Grant CEX2021-001131-S, both funded by MCIN/AEI/10.13039/501100011033. TS and VSD acknowledge financial support from the Spanish Ministry of Science, Innovation and Universities (MICIU) under grant PID2020-114822GB-I00.

Based on observations with the NASA/ESA/CSA {\em James Webb Space Telescope} obtained from the Barbara A. Mikulski Archive for Space Telescopes at the Space Telescope Science Institute (STScI), which is operated by the Association of Universities for Research in Astronomy, Incorporated, under NASA contract NAS5-03127. Support for Program number 1666 \citep{jwst1666} was provided through a grant from the STScI under NASA contract NAS5-03127.

The design and construction of HiPERCAM was funded by the European Research Council under the European Union’s Seventh Framework Programme (FP/2007-2013) under ERC-2013-ADG grant agreement number 340040 (HiPERCAM). VSD, SPL and HiPERCAM operations are funded by the Science and Technology Facilities Council (grant ST/Z000033/1). 

This work made use of Astropy:\footnote{http://www.astropy.org} a community-developed core Python package and an ecosystem of tools and resources for astronomy \citep{Astropy-2013, Astropy-2018, Astropy-2022}. This research made use of Photutils, an Astropy package for detection and photometry of astronomical sources. \citep{Bradley-2023}.

\section*{Data}
All the {{\em JWST} data used in this paper can be found in MAST: \dataset[10.17909/4z21-zv42]{http://dx.doi.org/10.17909/4z21-zv42}.

\bibliography{references.bib}{}
\bibliographystyle{aasjournal}



\appendix
\restartappendixnumbering

\section{Raw Light Curves}

\begin{figure*}
    \includegraphics[width=\textwidth]{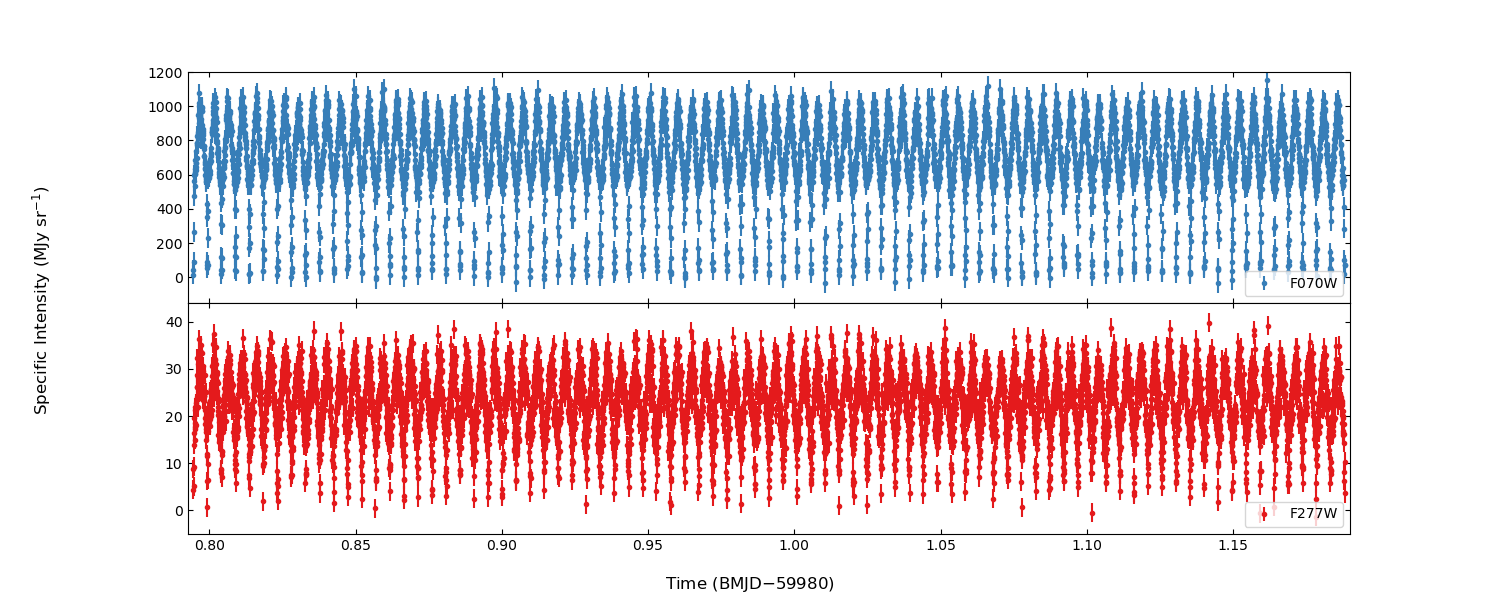}
    \caption{{\em JWST}/NIRCam Epoch 1 light curves of ZTF\,J1539 in the F070W (top) and F277W (bottom) bands.}
    \label{fig:E1_raw_lc}
\end{figure*}

\begin{figure*}
    \includegraphics[width=\textwidth]{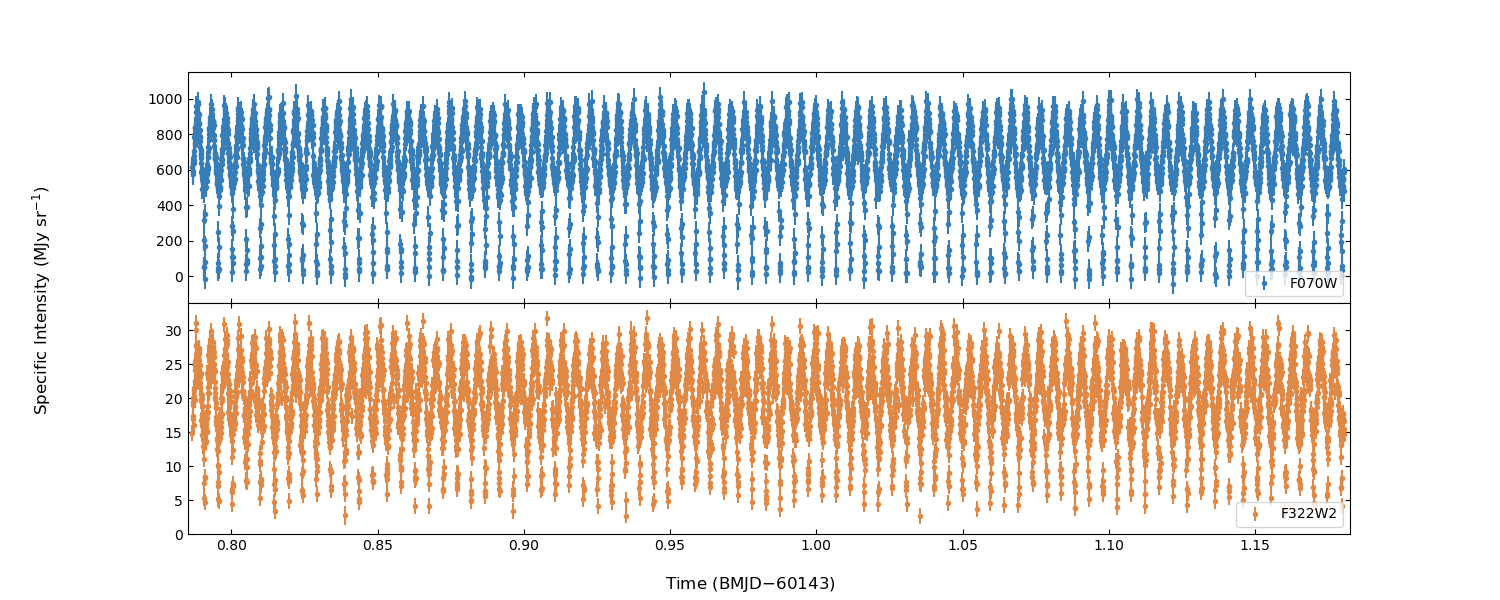}
    \caption{{\em JWST}/NIRCam Epoch 2 light curves of ZTF\,J1539 in the F070W (top) and F322W2 (bottom) bands.}
    \label{fig:E2_raw_lc}
\end{figure*}

\section{MCMC Corner Plots}

\begin{figure*}
    \centering
    \includegraphics[width=\textwidth]{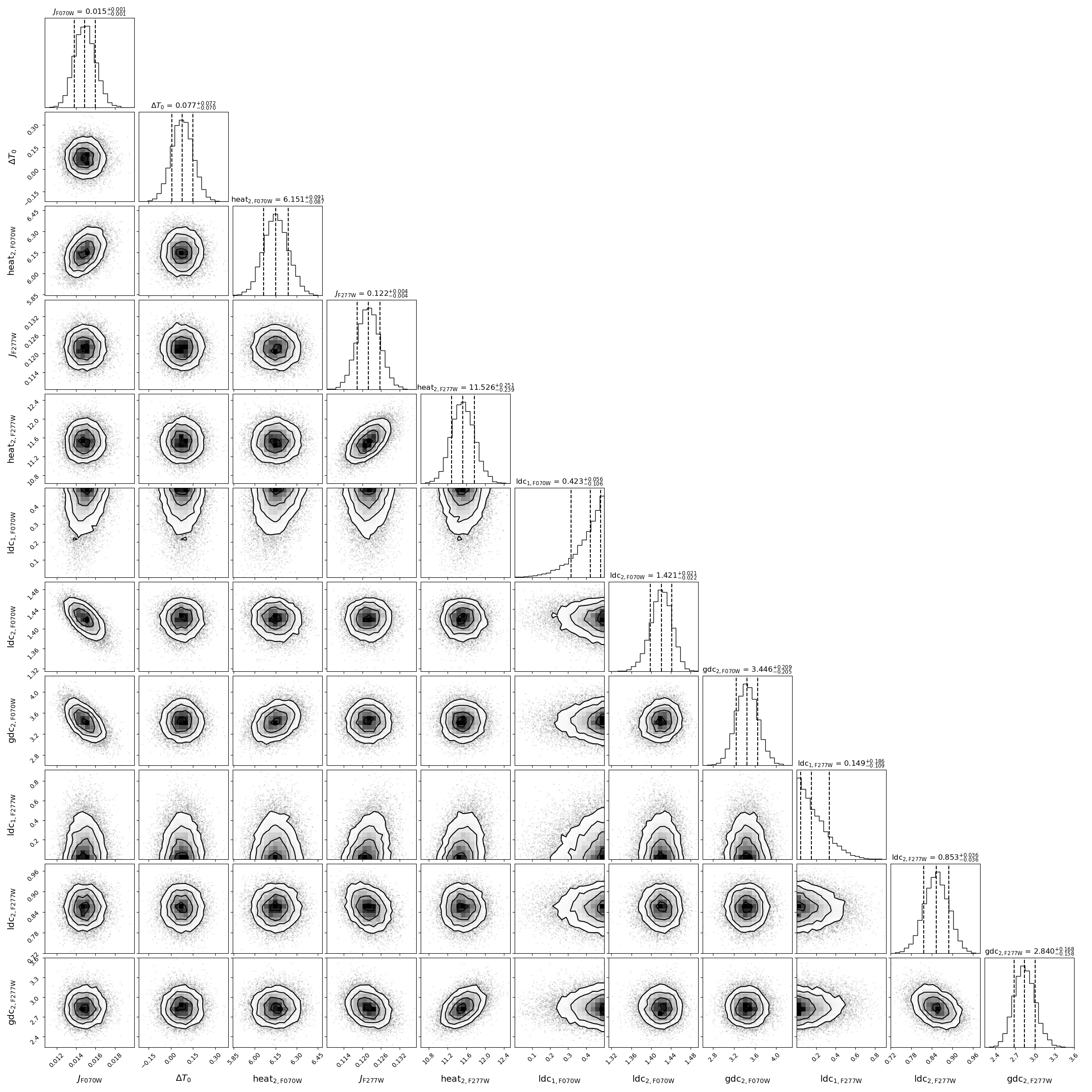}
    \caption{Corner plots of the fit to the light curve data from Epoch 1. Parameter definitions are given in Section \ref{sec:timing} and Table \ref{tab:final_par}.}
    \label{fig:E1_corner}
\end{figure*}

\begin{figure*}
    \centering
    \includegraphics[width=\textwidth]{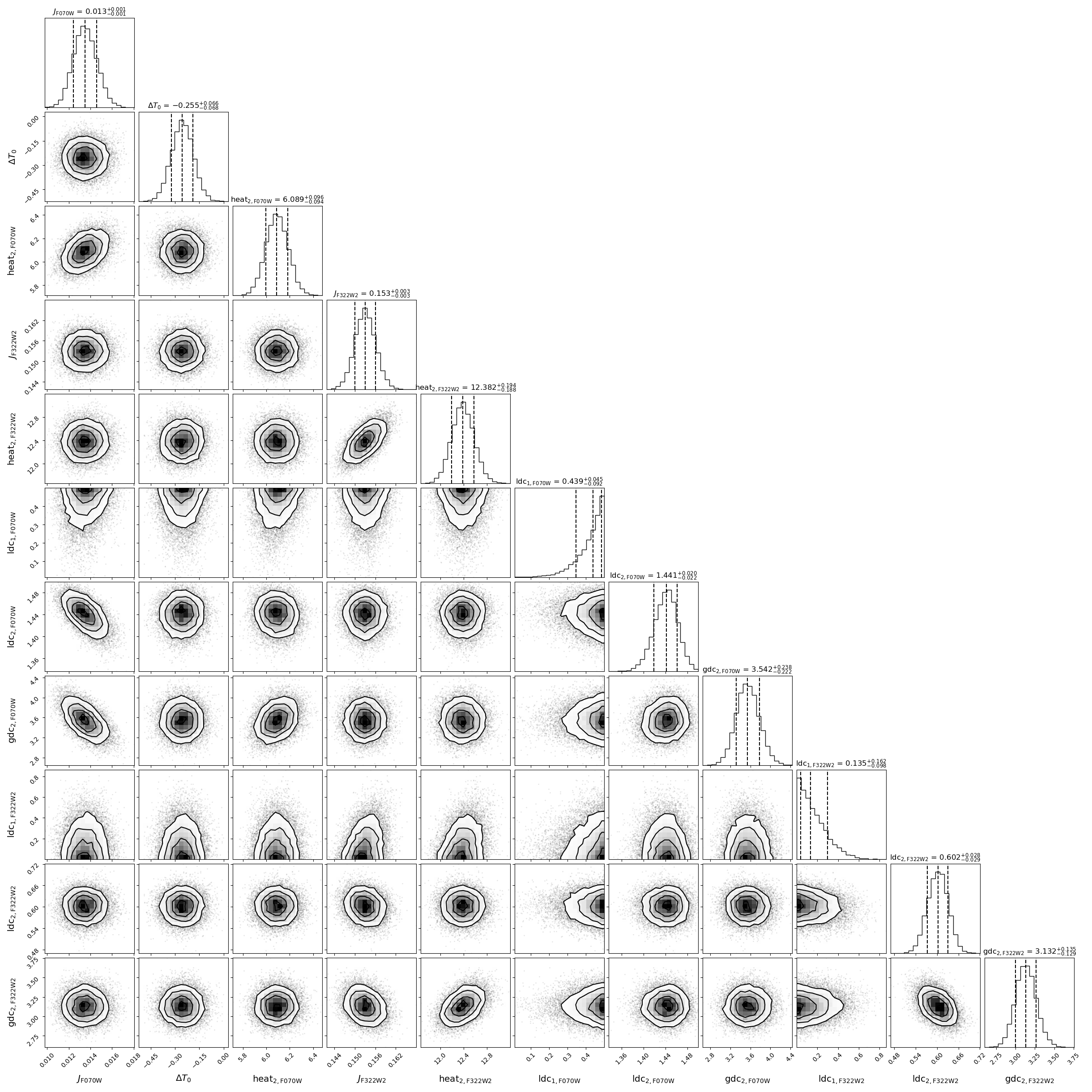}
    \caption{Corner plots of the fit to the light curve data from Epoch 2. Parameter definitions are given in Section \ref{sec:timing} and Table \ref{tab:final_par}.}
    \label{fig:E2_corner}
\end{figure*}

\begin{figure*}
    \centering
    \includegraphics[width=\textwidth]{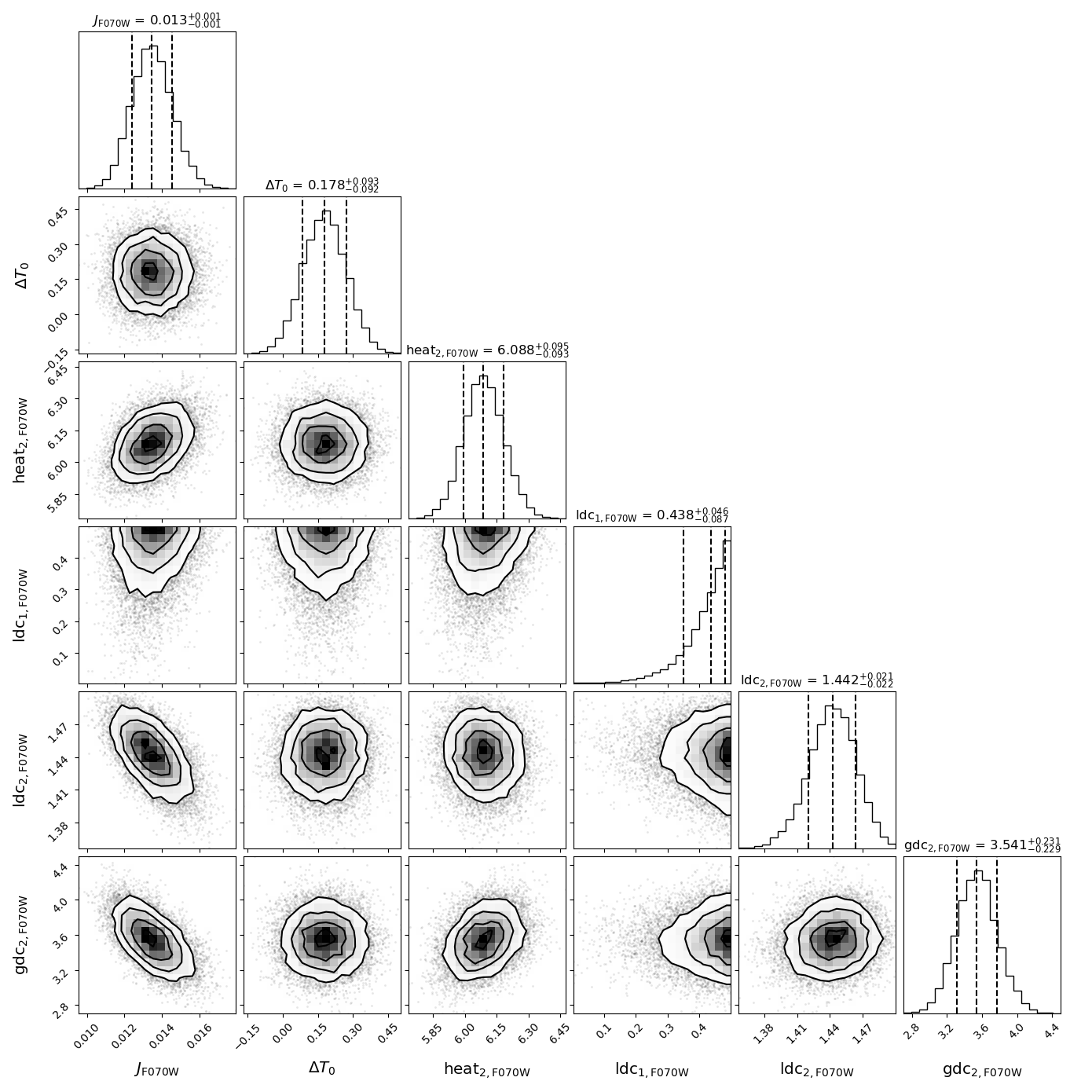}
    \caption{Corner plots of the fit to the F070W light curve in Epoch 2 after disregarding the F322W2 light curve from the fit.  Parameter definitions are given in Section \ref{sec:timing} and Table \ref{tab:final_par}.}
    \label{fig:E2_corner_SW}
\end{figure*}

\end{document}